\documentclass[preprint, amsmath, amssymb, floatfix, prb]{revtex4} 
\usepackage{graphicx}
\usepackage{dcolumn}
\usepackage{bm}

\begin{document}
\title{Dynamics of Macroscopic Wave Packet
Passing through \\ Double Slits: Role of Gravity and Nonlinearity}

\author{Katsuhiro~Nakamura}
\email{nakamura@a-phys.eng.osaka-cu.ac.jp}
\author{Naofumi~Nakazono}
\affiliation{Department of Applied Physics,
Osaka City University, Sumiyoshi-ku,
Osaka 558-8585 Japan}
\author{Taro~Ando}

\affiliation{Hamamatsu Photonics K.K.,  
Central Research Laboratory, Hirakuchi, Hamamatsu-City, Shizuoka, 
434-8601, Japan}

\date{June 20, 2005; revised July 19, 2005}

\begin{abstract}
Using the nonlinear Schr{\"o}dinger equation (Gross-Pitaevskii equation), 
the dynamics of a macroscopic wave packet for Bose-Einstein condensates
falling through double slits is analyzed. This problem is identified with 
a search for the fate of a soliton showing a head-on collision with a 
hard-walled obstacle of finite size. We explore the splitting of the wave 
packet and its reorganization to form an interference pattern. Particular 
attention is paid to the role of gravity ($g$) and repulsive nonlinearity 
($u_0$) in the fringe pattern. The peak-to-peak distance in the fringe 
pattern and the number of interference peaks are found to be proportional 
to $g^{-1/2}$ and $u_0^{1/2}g^{1/4}$, respectively. We suggest a way of 
designing an experiment under controlled gravity and nonlinearity.
\end{abstract}

\newcommand{\obs}[1]{\ensuremath{\overrightarrow{\boldsymbol{#1}}}}
\newcommand{\bol}[1]{\ensuremath{\boldsymbol{#1}}}
\newcommand{\suu}{\ensuremath{|\uparrow\uparrow\rangle}}
\newcommand{\sud}{\ensuremath{|\uparrow\downarrow\rangle}}
\newcommand{\sdu}{\ensuremath{|\downarrow\uparrow\rangle}}
\newcommand{\sdd}{\ensuremath{|\downarrow\downarrow\rangle}}
\newcommand{\suuu}{\ensuremath{|\uparrow\uparrow\uparrow\rangle}}
\newcommand{\suud}{\ensuremath{|\uparrow\uparrow\downarrow\rangle}}
\newcommand{\sudu}{\ensuremath{|\uparrow\downarrow\uparrow\rangle}}
\newcommand{\sudd}{\ensuremath{|\uparrow\downarrow\downarrow\rangle}}
\newcommand{\sduu}{\ensuremath{|\downarrow\uparrow\uparrow\rangle}}
\newcommand{\sdud}{\ensuremath{|\downarrow\uparrow\downarrow\rangle}}
\newcommand{\sddu}{\ensuremath{|\downarrow\downarrow\uparrow\rangle}}
\newcommand{\sddd}{\ensuremath{|\downarrow\downarrow\downarrow\rangle}}
\newcommand{\bsuuu}{\ensuremath{\langle\uparrow\uparrow\uparrow|}}
\newcommand{\bsuud}{\ensuremath{\langle\uparrow\uparrow\downarrow|}}
\newcommand{\bsudu}{\ensuremath{\langle\uparrow\downarrow\uparrow|}}
\newcommand{\bsudd}{\ensuremath{\langle\uparrow\downarrow\downarrow|}}
\newcommand{\bsduu}{\ensuremath{\langle\downarrow\uparrow\uparrow|}}
\newcommand{\bsdud}{\ensuremath{\langle\downarrow\uparrow\downarrow|}}
\newcommand{\bsddu}{\ensuremath{\langle\downarrow\downarrow\uparrow|}}
\newcommand{\bsddd}{\ensuremath{\langle\downarrow\downarrow\downarrow|}}
\newcommand{\balpha}{\boldsymbol{\alpha}}
\newcommand{\bbeta}{\boldsymbol{\beta}}
\newcommand{\bR}{\mathbf{R}}
\newcommand{\br}{\mathbf{r}}
\maketitle

%

Recently a great number of theoretical and experimental effort has been 
devoted to Bose-Einstein condensates (BECs). 
The trapping techniques can be used to control the typical size of localized 
states, such as a Gaussian wave packet, and the technique of Feshbach resonance 
can allow both the strength and sign of the nonlinearity to be varied.

BECs show a variety of interference phenomena that reflect the matter-wave 
nature. 
Among them, the BEC analog of Young's interference experiments is the most 
noteworthy. 
As demonstrated by Andrews {\it et al.}, a pair of BEC wave packets separately 
located in double potential wells begins to diffuse after the release of the 
trap and produces interference fringes\cite{Andre-1, Andre-2, Andre-3}. 
Since this pioneering experiment, numerous studies on interference have been 
devoted to creating a stable double-well trap to control the relative phase of 
split condensates\cite{Steng,Pitaev,Liu,Hind,Hans,Shin,Coll}, except for some 
novel experiments based on an atom-based Mach-Zehnder interferometer that 
exploits a superposition of internal states of BEC\cite{Kozu,Hagl,Torii}.

On the contrary, however, no work has been devoted to the BEC analog of single 
electron (photon) dynamics through real double slits (DSs). 
The release of a double-well trap cannot precisely mimic the passage through DSs. 
The former treats the second-order interference between two wave packets (WPs) 
with initially independent phases, while the latter treats the first-order 
interference between two fractions of the initially identical WP. 
A single BEC wave packet falling through DSs under uniform gravity will split 
into distorted pieces. 
However the fate of these split pieces is not clear because of the complicated 
effect of diffraction at the slit openings (windows). 
Furthermore, the effects of gravity and nonlinearity on double-slit interference 
have not been examined quantitatively up to now.

Both trapped and developing BECs are described using the Gross-Pitaevskii 
equation (GPE) or the nonlinear Schr{\"o}dinger equation 
(NSE)\cite{Pitaev, Liu, Coll, Holl, Roh}.
In this letter, by using NSE (GPE), the dynamics of a macroscopic WP falling 
through the DSs is analyzed.
In contrast to the ordinary quantum mechanics, BEC has advantage in that it shows 
a continuous time evolution of WP dynamics without being affected by the subtle 
problem of quantum demolition measurement.
We here explore how WP splitting occurs at the slit openings and how the split 
pieces reorganize to form the interference pattern.
We pay special attention to the effect of gravity and nonlinearity on the fringe
pattern.
We also discuss the designing of the experiment.

At low temperatures, $N_{0}$ atoms occupy the single quantum state and the 
many-body wave function is described by a single macroscopic wave function $\phi$ 
which satisfies NSE, i.e., the time-dependent Schr{\"o}dinger equation with the 
additional mean-field type nonlinear term. 
In the presence of a harmonic trap and gravity, GPE (NSE) becomes
\begin{equation}
i\hbar \frac{\partial \phi}{\partial t}=
-\frac{\hbar^{2}}{2m}\nabla^{2}\phi+V_{trap}({\bf r})\phi+
U_{gravity}({\bf r})\phi+
u_{0}|\phi|^{2}\phi ,
\label{eq:2}
\end{equation}
where $u_{0}$ is the nonlinear interaction defined by 
$u_{0}=\frac{4 \pi \hbar^{2} a}{ml}N_{0}$ [in 2-dimensional (2D) space] or 
$\frac{4 \pi \hbar^{2} a}{m}N_{0}$ (in 3D space) with $a$, $m$ and $l$ being 
scattering length, atomic mass and the characteristic length (to be defined below), 
respectively. 
$V_{trap}({\bf r})$ and $U_{gravity}({\bf r})$ are potentials for a harmonic trap 
and gravity, respectively.

Practically, a BEC WP in a harmonic trap exhibits a cigar-like 
shape.\cite{Andre-1, Andre-2, Andre-3} 
When the WP lengthens in the $z$ direction to form a nearly cylindrical shape, 
cross-sectional dynamics of the WP, which is observed on the $x$-$y$ plane 
bisecting the $z$ axis of the WP, can be approximately described using a 2D GPE 
without losing any physical meaning. 
Thus we perform 2D analysis that requires less computational cost than the 3D 
version.
We assume an initial circularly trapped WP that falls toward the positive $y$ 
direction under uniform gravity (gravity constant $g$). 
Then we set 
$\nabla^{2}=\frac{\partial ^{2}}{\partial x ^{2}}
 +\frac{\partial^{2}}{\partial y ^{2}}$ 
and
\begin{equation}
V_{trap}({\bf r})=\frac{1}{2}m\omega ^{2}(x^{2}+y^{2} ); \qquad
U_{gravity}({\bf r})=-mgy.
\label{eq:3}
\end{equation}
Using the characteristic length
$l=\sqrt{\hbar/(m\omega)}$, one can make all variables dimensionless
as
\begin{equation}
t^{\prime}=\omega t, \,  x^{\prime}=\frac{x}{l}, \, y^{\prime}=\frac{y}{l}, \,
\phi^{\prime}=l\phi .
\label{eq:4a}
\end{equation}
Then GPE is reduced to 
\begin{equation}
i\frac{\partial \phi^{\prime}}{\partial t^{\prime}}=
-\frac{1}{2} \left( \frac{\partial ^{2}}{\partial x^{\prime 2}}+
\frac{\partial^{2}}{\partial y^{\prime 2}} \right) \phi^{\prime}
+\frac{1}{2}(x^{\prime 2}+y^{\prime 2})\phi^{\prime} 
 -g^{\prime}y^{\prime}\phi^{\prime}
 +u_{0}^{\prime }|\phi^{\prime}|^{2}\phi^{\prime}, 
\label{eq:4}
\end{equation}
where $u_{0}^{\prime }=4\pi N_{0}a/l$ and
$g^{\prime}=g/(l\omega^{2})$ are also dimensionless constants.
Initially, we ignore the contribution from gravity and, using the confining 
potential, prepare a circularly symmetric Gaussian WP with its center of mass
at the origin ${\bf r}=(0,0)$,
\begin{equation}
\phi^{\prime}_{ini}=\frac{1}{ \sqrt{\pi} 
  \left( 1+\frac{u_0^\prime}{2\pi} \right)^{1/4}}
\exp \left(-\frac{x^{\prime 2}+y^{\prime 2}}{2 \sqrt{1+\frac{u_{0}^{\prime
}}{2\pi}}}\right) ,
\label{eq:5}
\end{equation}
which minimizes the total energy 
$E^{\prime}=\int d^{2} {\bf r}^{\prime} ( \frac{1}{2}|\nabla
\phi^{\prime}|^{2} + V_{trap}^{\prime}|\phi^{\prime}|^{2}
 + \frac{u_{0}^{\prime}}{2}|\phi^{\prime}|^{4} )$.
Here, the de Broglie wavelength of an initially stationary WP can be longer 
than the size of the WP itself. 
Thus small variation of the WP shape contributes little to the global phenomena. 
The Gaussian WP profile is a simple substitute for an exact solution of GPE 
as far as short-time dynamics of the WP is concerned.

Our strategy is as follows: 
(1) release the confining potential and switch on the uniform gravity;
(2) WP falls on DS;
(3) WP splits as it passes through the DS;
(4) the split pieces reorganize to form a fringe pattern.
Dynamics of the WP is obeyed by GPE in eq.(\ref{eq:4}) together with the Dirichlet 
boundary condition at DS: hard walls forming the DS are taken to be one dimensional 
along the $x$ direction.
We assume the interference phenomena appear in the short-time dynamics shorter than
the decoherence time and use GPE even after releasing the trap, closely following 
the procedures in most of the existing works\cite{Pitaev, Liu, Coll, Holl, Roh}.

Given any arbitrary value $u_{0}^{\prime}$, we systematically investigate the effect 
of gravity on the fringe pattern, starting from a common WP with fixed width.
For this purpose, we always choose WP in the case of vanishing nonlinearity 
($u_{0}^{\prime} = 0$),
\begin{equation}
\phi^{\prime }_{ini}=\frac{1}{\sqrt{\pi}}
\exp \left[ -\frac{1}{2}(x^{\prime 2}+y^{\prime 2}) \right] , 
\label{eq:12}
\end{equation}
as an initial state for any value of nonlinearity.
This means that we have replaced $V_{trap}$ in eq.(\ref{eq:4}) by the 
$u_{0}^{\prime }$-dependent confining potential, 
$\frac{1}{2}(1+\frac{u_{0}^{\prime}}{2\pi})(x^{\prime 2}+y^{\prime 2})$, but have 
taken the scaling in eq.(\ref{eq:4a}) using $l$ proper for $V_{trap}$ with 
$u_{0}^{\prime}=0$.
Since we consider the falling of WP after releasing $V_{trap}$, the formal change 
of the confining potential has no effect on the dynamics under gravity.

\begin{figure}[t]
     \begin{minipage}{.37\textwidth}
       \includegraphics[width=8cm]{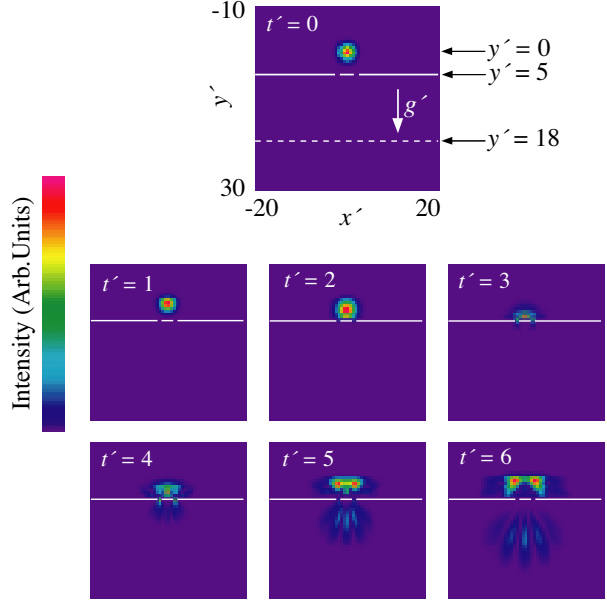}
          \end{minipage}
\caption{Wave packet dynamics with $u_{0}^{\prime}=0$, $g^{\prime}=1$.
Intensity distributions of WP at $t^{\prime}=0.0$ and $t^{\prime}=$1.0--6.0.
Topmost panel illustrates the setting for double-slits
experiments.}
\label{linearcase}
\end{figure}

We numerically integrate GPE in three ways, i.e., alternating direction implicit 
method (ADI method), Crank-Nicholson's method, and split-step method, all of which 
are found to yield identical results.
Figure~\ref{linearcase} shows the $x^{\prime}$-$y^{\prime}$ plane of size 
$40{\times}40$ that we employ.
Units for spatial and temporal grids are
$\delta x^{\prime}=\delta y^{\prime}=0.1$, $\delta t=0.002$.
The center of WP, being located at the origin at $t=0$, begins to fall along the 
$y^{\prime}$ direction due to the uniform gravity.
DS is set at $y^{\prime}=5$. Lengths for the central stopper and a pair of windows 
are 3 and 1, respectively. 
The fringe pattern will be observed on a virtual screen located at 
$y^{\prime}=L_{sc}=18$.
The final time of the present computation is the time when the classical particle 
(center of mass) reaches the screen, and is given by
$t=T_{max}=\sqrt{2L_{sc}/g^{\prime}}$.
Up to this time, we detect little fraction of WP escaping outside the whole grid 
area through the boundaries at $x^{\prime}=\pm 20$.

The time evolution of WP in the linear case ($u_{0}^{\prime}=0$) is given in 
Fig.~\ref{linearcase}.
The gravity is fixed at $g^{\prime}=1$.
Figure~\ref{linearcase} shows that the falling WP breaks into a pair of pieces 
during passage through the DS and then begins to form a clear fringe pattern.
Above the DS we find a remnant of WP, which has failed to fall through the DS, 
interferes parallel to DS. Because of the gravity, the remnant will also fall
through DS sooner or later in due course.
Our additional calculation (to be described elsewhere) indicates that corresponding 
WP dynamics in the case when either one of the DS windows is closed leads to no 
fringe pattern and merely to a single humped diffusive WP.
To see the fringe pattern, therefore, it is essential to open both windows.
The results are consistent with the result of a single electron interference using 
a DS.
\begin{figure}[h]
\begin{minipage}{.37\textwidth}
\includegraphics[width=8cm]{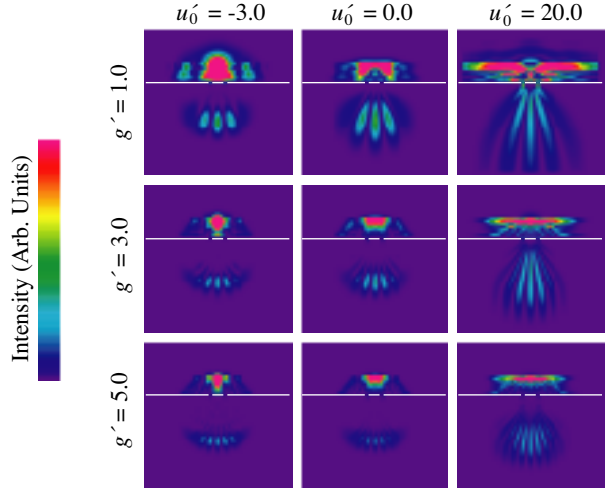}
\end{minipage}
\caption{Intensity distribution of WP.
   From top to bottom: $g^{\prime}=$1, 3, and 5.
   From left to right: $u_{0}^{\prime}=-3$, 0, and 20.}
\label{nineshots}
\end{figure}

\begin{figure}
\begin{minipage}{.37\textwidth}
\includegraphics[width=8cm]{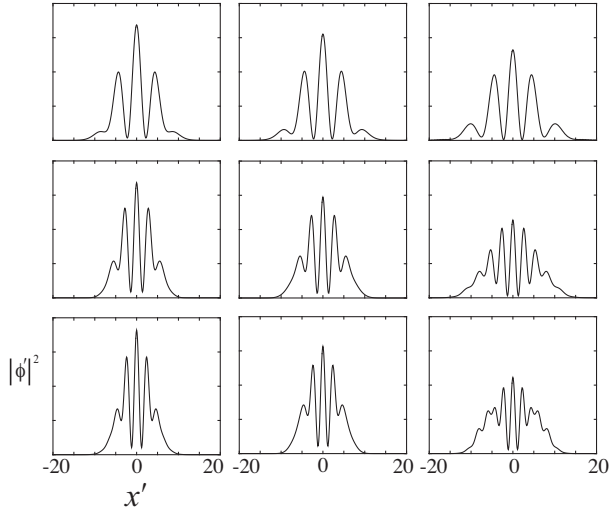}
\end{minipage}
\caption{Fringe pattern on screen. Probability ($0-0.2$)
versus horizontal coordinate. Arrangement of panels is the same as in
Fig.~\ref{nineshots}.}
\label{ninesections}
\end{figure}

Taking the result for the linear case and $g^{\prime}=1$ as a standard pattern, let 
us proceed to the investigation of the role of tunable nonlinearity and gravity.
We shall focus on the fringe pattern on the screen at $y^{\prime}=L_{sc}$ at 
$t=T_{max}$.
We choose $u_{0}^{\prime}=-3$, 0, 20 for the nonlinear term and $g^{\prime}=$1, 3, 5 
for gravity.
Figure~\ref{nineshots}, which includes nine fringe patterns, represents the intensity 
distribution of WP at $t=T_{max}$, and Fig.~\ref{ninesections} shows their cross 
sections on the screen, with the total probability amplitude for this section being 
normalized to unity.
We find that the nonlinearity is responsible for the localization or delocalization
of WP, while gravity governs the velocity of WP arriving at the screen and is 
responsible for the de Broglie wavelength.
For a fixed nonlinearity, the increased gravity will shorten the de Broglie wavelength 
at time $t=T_{max}$. 
On the other hand, for a fixed gravity, the increased nonlinearity
$u_{0}^{\prime}$ will broaden the fringe pattern.

\begin{figure}[t]
\begin{minipage}{.38\textwidth}
\begin{center}
\includegraphics[width=7.5cm]{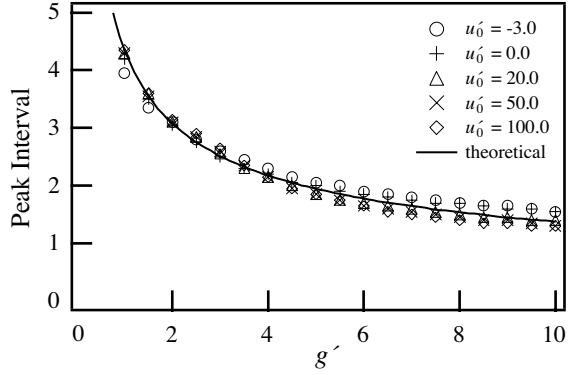}
\end{center}
\end{minipage}
\caption{Main interval versus gravity}
\label{peakinterval}
\end{figure}

Let us develop a theoretical insight in greater detail.
Under a fixed nonlinearity, by using Fig.~\ref{ninesections}, one can evaluate the 
peak-to-peak interval.
The gravity dependence of the main interval (between the central and second peaks) 
is shown in Fig.~\ref{peakinterval}.
For each value of $u_{0}^{\prime}$, the interval decreases monotonically as a 
function of $g^{\prime}$.
The peak interval is proportional to the de Broglie wavelength defined by
$\lambda ^{\prime} \equiv 2\pi/p^{\prime}$. 
If we choose $p^{\prime}$ to be the momentum for the classical particle corresponding 
to the center of WP, 
$p^{\prime}=V_{max}=g^{\prime}T_{max} = g^{\prime}\sqrt{2L_{sc}/g^{\prime}}=\sqrt
{2g^{\prime}L_{sc}}$. 
Then we obtain 
\begin{equation}
\lambda^{\prime} =\frac{2\pi}{\sqrt{2g^{\prime}L_{sc}}} 
\end{equation}
regardless of the value $u_{0}^{\prime}$, the validity of which is confirmed in 
Fig.~\ref{peakinterval}.

On the other hand, the nonlinearity is responsible for WP broadening, namely, 
the extension of the area in which a fringe pattern can be formed.
We shall make a simple evaluation of WP width at $t=T_{max}$.
Let us approximate the falling WP by a cylinder ($\phi = \pi^{-1/2}R^{-1}$) with 
radius $R$.
By suppressing gravity, GPE becomes circularly symmetric and is given by
\begin{equation}
i\dot{\phi}=\left( -\frac{\partial^{2}}{2\partial
R^{2}}-\frac{1}{2R}\frac{\partial}{\partial R}+
u_{0}^{\prime}|\phi|^{2} \right) \phi.
\end{equation}
Using the above cylindrical solution in this equation, we have
$\dot{R}=i(u_{0}^{\prime}-\pi/2)\pi^{-1}R^{-1}$, whose solution is
$R\approx \sqrt{(u_{0}^{\prime}-\pi/2)t}$ in the large $t$ region.
Therefore, at $t=T_{max}$,
\begin{equation}
R_{sc} \propto \sqrt{\frac{u_{0}^{\prime}-\pi/2}{\sqrt{g^{\prime}}}} .
\end{equation}

We then proceed to evaluate the peak number. 
The peak-to-peak interval $\Delta x^{\prime}_{peak}$ is approximately equal to 
the de Broglie wavelength $\lambda^{\prime}$ at  $t=T_{max}$, so 
$\Delta x^{\prime}_{peak}\propto \lambda^{\prime} \propto 1/\sqrt{g^{\prime}}$.
Eventually, the peak number (PN) at the screen can be given by
\begin{equation}
{\rm PN}=\frac{R_{sc}}{\Delta x^{\prime}_{peak}}\propto
\sqrt{\left( u_{0}^{\prime}-\frac{\pi}{2} \right)\sqrt{g^{\prime}}} .
\label{eq:19}
\end{equation}

This approximate analytical expression for PN explains how the peak number
depends on the nonlinearity and gravity.
Table~\ref{NumberTable} obtained from our numerical fringe pattern on the screen,
lists PN for various values of $g^{\prime}$ and $u_{0}^{\prime}$ and is consistent 
with eq.(\ref{eq:19}).
\begin{table}
\begin{center}
\begin{minipage}{.37\textwidth}
\includegraphics[width=6cm]{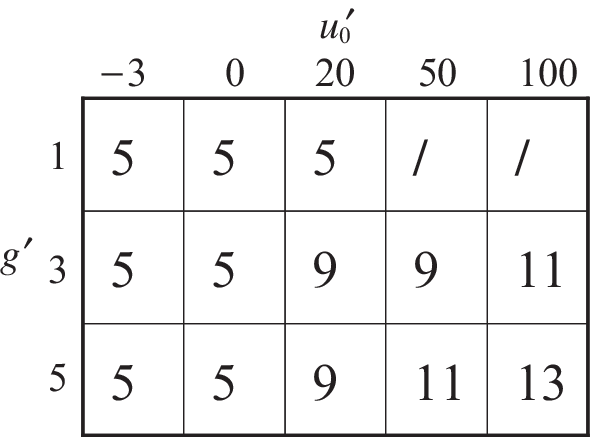}
\end{minipage}
\end{center}
\caption{Peak number}
\label{NumberTable}
\end{table}

The transmission probability of WP at $t=T_{max}$ is the portion of WP that lies 
below the DS at that time.
This probability is affected by the nonlinearity.
Let us define ${\rm Pr}(A)$ and ${\rm Pr}(B)$
for each portion lying above and below DS, respectively.
They are calculated at $t=T_{max}$ as
\begin{equation}
{\rm Pr}(A \; {\rm or}\;  B)=\int \! \! \int_{A \; {\rm or} \; B} 
  dx^{\prime}dy^{\prime} |\phi^{\prime}|^{2} .
\label{eq:20}
\end{equation}
Noting ${\rm Pr}(A)+{\rm Pr}(B)=1$, ${\rm Pr}(B)$ gives the transmission probability.
${\rm Pr}(B)$ at $t=T_{max}$ is found to depend on $u_{0}^{\prime},g^{\prime}$ and 
takes values in the range $0.1\leq {\rm Pr}(B)\leq 0.3$.
For larger $u_{0}^{\prime}$, WP broadens before reaching the DS and therefore 
${\rm Pr}(B)$ diminishes. 
On the other hand, for increased $g^{\prime}$, WP reaches the DS without broadening 
very much because of the highly accelerated motion, and a large portion of WP dwells 
around the central stopper, which suppresses ${\rm Pr}(B)$. 
Thus ${\rm Pr}(B)$ is sensitive to the values of $u_{0}^{\prime}$ and $g^{\prime}$.
However, qualitative features of the interference, such as $g^{\prime}$ and 
$u_{0}^{\prime}$ dependences of the peak-to-peak distance and peak number, remain 
unchanged.

We have so far investigated WP dynamics using scaled time and coordinates.
Here we suggest an experimental situation to which the present result is applicable.
Let us introduce the material-specific constant $\alpha=\sqrt{\hbar/m}$.
The value $g^{\prime}$ that we have employed determines $\omega$ and $l$.
With the use of $l=\sqrt{\hbar/(m\omega)}$ and
$g^{\prime}=g/(l\omega^{2})$, one obtains
\begin{equation}
\omega =\left( \sqrt{\frac{m}{\hbar}}\frac{g}{g^{\prime}} \right)^{2/3}=
\left( \frac{1}{\alpha} \right)^{2/3} \left( \frac{g}{g^{\prime}} \right)^{2/3}
\label{eq:22}
\end{equation}
and
\begin{equation}
l=\alpha^{4/3}\left( \frac{g}{g^{\prime}} \right)^{-1/3}.
\label{eq:23}
\end{equation}
The last value represents the scale of, for example, the WP width, the DS, and the 
stopper. 
Let us consider more concretely the case of $^{87}{\rm Rb}$ and $g^{\prime}=1$.
Noting that $\alpha=2.726\times 10^{-5}$ and the gravity constant on the earth 
$g=9.80665$~m/s, we obtain $l=0.383$~$\mu$m.
In a laboratory of the space station, one can expect a weak  gravity, e.g., 
$\hat{g}=g/1000$. 
Then the experiment corresponding to $g^{\prime}=1$ is realized on a length scale 
$\hat{l}$ that satisfies
\begin{equation}
\frac{\hat{l}}{l}= \left( \frac{\hat{g}}{g} \right)^{-1/3}=10,
\end{equation}
from which we obtain $\hat{l}=3.83$~$\mu$m.
The above equality indicates that on decreasing the gravity scale by three orders,
the corresponding length scale increases by one order. 
In other words, the length scale is almost insensitive to the change of gravity and 
the size of DS can be kept almost constant when designing the present experiment 
under a weak gravitational field.

We have explored the effect of nonlinearity and gravity on the macroscopic fringe 
pattern generated by BEC passing through DS. 
Most of the activities so far concerning the BEC analog of Young's interference 
experiment referred to a double-well trap leading to the 2nd-order interference 
between two initially separated WPs with uncontrolled independent phases.
By contrast we mimic the BEC analog of the 1st-order interference (a single electron 
interference).
A study of the BEC dynamics through the DS is identified with a search for the fate 
of a 2D soliton showing a head-on collision with a hard-walled obstacle of finite size.
In our simulation, we find that split pieces of WP below the DS successfully reorganize 
to form a clear fringe pattern.
The analytic expressions are derived for the peak-to-peak distance and peak number as 
a function of gravity and nonlinearity.
It should be noted that our preliminary simulation for a more realistic 3D version of 
the present system yielded the same result as found here.
A further analysis of wave-function patterns, such as nodal lines and phase rigidity,
which is developed in the context of quantum chaos\cite{berg}, are subjects which we 
intend to study in the future.

K. N. is grateful to M. Ueda, M. Hosoda, M. Wilkinson and C. Kosaka for enlightening 
discussions. T. A. thanks M. Mizobuchi and Y. Ohtake for discussions on the experimental 
aspects of BECs.


\end{document}